\documentclass[sigconf,screen]{acmart}
\usepackage[utf8]{inputenc}
\usepackage{booktabs}
\usepackage[utf8]{inputenc}
\usepackage{lipsum}
\usepackage{graphicx}
\usepackage{subcaption}
\usepackage{array}
\usepackage{balance}
\usepackage{flushend}
\definecolor{lightgray}{gray}{0.92}

\newcommand{\overbar}[1]{\mkern 1.5mu\overline{\mkern-1.5mu#1\mkern-1.5mu}\mkern 1.5mu}

\title[Characterizing the Transition to Kotlin of Android Apps: A Study on F-Droid, Play Store, and GitHub]{Characterizing the Transition to Kotlin of Android Apps: \\A Study on F-Droid, Play Store, and GitHub}

\setcopyright{acmcopyright}
\acmPrice{15.00}
\acmDOI{10.1145/3340496.3342759}
\acmYear{2019}
\copyrightyear{2019}
\acmISBN{978-1-4503-6858-2/19/08}
\acmConference[WAMA '19]{Proceedings of the 3rd ACM SIGSOFT International Workshop on App Market Analytics}{August 27, 2019}{Tallinn, Estonia}
\acmBooktitle{Proceedings of the 3rd ACM SIGSOFT International Workshop on App Market Analytics (WAMA '19), August 27, 2019, Tallinn, Estonia}

\begin{document}

\author{Riccardo Coppola}
\affiliation{%
\institution{Politecnico di Torino}
\streetaddress{Corso Duca degli Abruzzi 24}
\city{Turin}
\state{Italy}
\postcode{10129}
}
\email{riccardo.coppola@polito.it} 
\author{Luca Ardito}
\affiliation{%
\institution{Politecnico di Torino}
\streetaddress{Corso Duca degli Abruzzi 24}
\city{Turin}
\state{Italy}
\postcode{10129}
}
\email{luca.ardito@polito.it} 
\author{Marco Torchiano}
\affiliation{%
\institution{Politecnico di Torino}
\streetaddress{Corso Duca degli Abruzzi 24}
\city{Turin}
\state{Italy}
\postcode{10129}
}
\email{marco.torchiano@polito.it}

\begin{abstract}
Context: Kotlin is a novel language that represents an alternative to Java, and has been recently adopted as a first-class programming language for Android applications. Kotlin is achieving a significant diffusion among developers, and several studies have highlighted various advantages of the language when compared to Java.

Goal: The objective of this paper is to analyze a set of open-source Android apps, to evaluate their transition to the Kotlin programming language throughout their lifespan and understand whether the adoption of Kotlin has impacts on the success of Android apps.

Methods: We mined all the projects from the F-Droid repository of Android open-source applications, and we found the corresponding projects on the official Google Play Store and on the GitHub platform. We defined a set of eight metrics to quantify the relevance of Kotlin code in the latest update and through all releases of an application. Then, we statistically analyzed the correlation between the presence of Kotlin code in a project and popularity metrics mined from the platforms where the apps were released.

Results: Of a set of 1232 projects that were updated after October 2017, near 20\% adopted Kotlin and about 12\% had more Kotlin code than Java; most of the projects that adopted Kotlin quickly transitioned from Java to the new language. The projects featuring Kotlin had on average higher popularity metrics; a statistically significant correlation has been found between the presence of Kotlin and the number of stars on the GitHub repository.

Conclusion: The Kotlin language seems able to guarantee a seamless migration from Java for Android developers. With an inspection on a large set of open-source Android apps, we observed that the adoption of the Kotlin language is rapid (when compared to the average lifespan of an Android project) and seems to come at no cost in terms of popularity among the users and other developers. 

\end{abstract}

\begin{CCSXML}
<ccs2012>
<concept>
<concept_id>10011007</concept_id>
<concept_desc>Software and its engineering</concept_desc>
<concept_significance>500</concept_significance>
</concept>
<concept>
<concept_id>10011007.10011006.10011072</concept_id>
<concept_desc>Software and its engineering~Software libraries and repositories</concept_desc>
<concept_significance>300</concept_significance>
</concept>
<concept>
<concept_id>10011007.10011074.10011092</concept_id>
<concept_desc>Software and its engineering~Software development techniques</concept_desc>
<concept_significance>300</concept_significance>
</concept>
<concept>
<concept_id>10011007.10011074.10011111.10011113</concept_id>
<concept_desc>Software and its engineering~Software evolution</concept_desc>
<concept_significance>300</concept_significance>
</concept>
</ccs2012>
\end{CCSXML}

\ccsdesc[500]{Software and its engineering}
\ccsdesc[300]{Software and its engineering~Software libraries and repositories}
\ccsdesc[300]{Software and its engineering~Software development techniques}
\ccsdesc[300]{Software and its engineering~Software evolution}

\keywords{App Market Analytics, Mobile Development, Kotlin, Java, Software Maintenance, Empirical Software Engineering}

\maketitle

\section{Introduction}

The Android OS has established itself as the preferred operating system among mobile users, and as one of the most popular OSs overall (74.85\% of the mobile market share as of April 2019\footnote{http://gs.statcounter.com/os-market-share/mobile/worldwide}). Several app markets are available for Android developers to release or sell their apps, such as the official Google Play store, the Amazon AppStore, F-Droid, Aptoide, GetJar, itch.io. F-Droid\footnote{https://f-droid.org/} is a repository of free and open-source apps for Android devices, of which both the apk with the compiled code and a source tarball are provided. F-Droid maintains a brief track of the history of the released applications, allowing the users to download the latest three releases of any published app (as opposed to the Play Store, that keeps only the latest released .apk for each family of devices). Many of the applications released on F-Droid are modified versions of apps released to other markets by their developers \cite{7832927}, or initially open-source apps that have since their first release become closed source.

Since the beginning of Android programming, several development approaches and frameworks have been proposed \cite{martinez2017towards}. Kotlin is a new programming language, appeared in 2011, capable of running on the Java Virtual Machine; it represents an alternative to and can seamlessly coexist with Java. Kotlin is described as a safer, more concise alternative to Java \cite{kt1}, and among its selling points there is the possibility of avoiding several common Java development pitfalls such as Nullability, Mandatory Casts, Long argument lists and Data classes\footnote{https://kotlinlang.org/docs/kotlin-docs.pdf}. The first stable release of Kotlin was distributed in February 2016, and in May 2017 Kotlin became a first-class language on Android, with support provided by the Android Studio IDE since release 3.0 (October 2017). For these reasons, Kotlin is gaining traction with Android software developers.

This work wants to capture a snapshot of the current diffusion of Kotlin on the F-Droid repository, to evaluate the impact that the adoption of Kotlin code has on the user perception of the apps, and to analyze the history of evolution of the programming language during the release history of the apps released on such platform. To do so, we carried an experiment by mining all repositories on the platform and finding the correspondent ones on the Google Play Store and among the OS repositories hosted on GitHub. We defined a set of static metrics to quantify the amount of Kotlin code available on a generic source code package and to characterize the translation from Java to Kotlin, and applied statistic tests in order to check correlations between Kotlin adoption and popularity metrics.

The remainder of the present manuscript is organized as follows: section 2 summarizes the design of the study; section 3 reports the results and gives answers to the defined Research Questions; section 4 reports the Threats to the validity of the current study; section 5 reports the findings of related works available in literature; finally, section 6 discusses the implications of this paper and provides hints for future work.

\section{Study Design}

\begin{table}
\small
    \centering
        \caption{GQM Template for the study}

    \begin{tabular}{@{}r@{~:~}l@{}}
    \toprule
        Object of Study & Kotlin programming language\\
        %\midrule
        Purpose & Investigate  Kotlin presence in open-source projects\\
        %\midrule
        Focus & Diffusion, Evolution, Popularity Metrics\\
        %\midrule
        Context & Mobile applications released on F-Droid \\
        %\midrule
        Stakeholders & Developers, Researchers\\
        \bottomrule
    \end{tabular}
    \label{tab:gqm}
\end{table}

We report the design, goal, research questions, metrics and procedure adopted for the study following the Goal Question Metric (GQM) paradigm \cite{caldiera1994goal}, as summarized in table \ref{tab:gqm}.

The \emph{goal} of the study was to give a characterization of the migration of Android open-source projects to the Kotlin programming language, and to investigate the correlation between the presence of Kotlin and a set of metrics related to the app popularity. The study was based on a starting set of applications whose source code was released on the F-Droid platform. The results of the study are interpreted according to the perspective of \emph{developers} of Android Apps, as well as \emph{researchers} of the field.

\subsection{Research Questions and Metrics}

In this section we detail the Research Questions that we have defined to pursue the goal of the study. Since the goal of the study was primarily explorative, the first two RQs were purely descriptive.

The first objective of the study was to quantify the amount of apps on the OS repository F-Droid that featured Kotlin code, and their diffusion through time. We hence formulated the following research question:\smallskip

\textbf{RQ1 - Diffusion:} What is the adoption of Kotlin on Android apps available on F-Droid?\smallskip

To measure the diffusion of Kotlin applications and the relative importance of Kotlin code in Android projects, we defined the following metrics:

\begin{itemize}

\item\textbf{KRL} (Kotlin Relative LOCs), i.e. the number of Kotlin LOCs over the total amount of production LOCs of the project;

\item\textbf{KRF} (Kotlin Relative Files), i.e. the number of Kotlin .kt files over the total amount of production code files (.kt + .java files);

\item\textbf{KFPR} (Kotlin-Featuring Projects Ratio), i.e. the ratio of projects of a set featuring at least a Kotlin file;

\item\textbf{KMPR} (Kotlin-Majority Projects Ratio), i.e. the ratio of projects of a set featuring a majority of Kotlin LOCs in production code.

\end{itemize}\medskip

The second objective of our study was to inspect how and when the project featuring Kotlin migrated from Java to Kotlin during their lifespan. Hence, our second research question can be formulated as:\smallskip

\textbf{RQ2 - Evolution:} How have projects on F-Droid evolved from Java to Kotlin? \smallskip

To answer RQ2, we defined the following metrics: 

\begin{itemize}
    \item \textbf{KNR} (No-Kotlin Relative Releases), i.e. the ratio of tagged releases without Kotlin code;
    \item \textbf{KAR} (Kotlin Adoption relative Releases), i.e. the ratio of tagged releases that featured less than 50\% Kotlin code;
    \item \textbf{KMR} (Kotlin Majority relative Releases), i.e. the ratio of tagged releases that featured a majority of Kotlin LOCs;
    \item \textbf{KOR} (Only-Kotlin Relative Releases), i.e. the ratio of tagged releases that featured only Kotlin code.
\end{itemize}\medskip

The third objective of our study was to understand whether the usage of Kotlin had any effect on the popularity of the released app among its users or to other developers. Our third research question could hence be formulated as: \smallskip

\textbf{RQ3: Popularity - }Does the development with Kotlin have an influence on the success of released apps?\smallskip

To answer this question, we have sought for correlations between the diffusion metrics measured to answer RQ1, and popularity metrics that could be mined for projects released also on the Play Store or available on GitHub -- since no quality metric is available on the F-Droid platform. %Specifically, we based our analysis on the ratings and number of donwloads of applications on the Play Store, and on the number of stars of applications published on GitHub.

\subsection{Instruments}

\subsubsection{Mining of Packages from F-Droid} The first step of the procedure was a mining of all the projects and the related information from the F-Droid repository. 

%F-Droid is a repository of open-source Android application, as well as a marketplace that can be installed on the devices as an .apk to download os software. The presence of an application on F-Droid is not mutually exclusive with the presence of the Google Play Store. 

To mine projects from F-Droid we leveraged Selenium with Chromedriver\footnote{https://www.seleniumhq.org/}, creating a Java class able to mine all the information of the packages hosted on the platform, and to download the .tar.gz archive files containing their source code and the .apk distributable files. With such scraper, we were able to mine for each app the package name, the description, the last version (both the progressive version number registered on the F-Droid platform and the semantic version number assigned to the release by the developers), and the last date of update. The last scraping was performed as of May 17, 2019. %To ensure replicability of our experiment, we uploaded on our servers the .apk and .tar.gz files related to all the mined applications, since F-Droid maintains on their server only the three most recent updates for each hosted application, and hence the source code of the analyzed apps might me not available for reproductions of the study.

\subsubsection{Static Analysis of F-Droid Packages} For each source code package downloaded from F-Droid, we performed a static analysis to measure the Diffusion metrics. We created bash scripts to this purpose, leveraging the cloc tool\footnote{http://cloc.sourceforge.net/} for counting LOCs based on the language, and the rg tool\footnote{https://github.com/BurntSushi/ripgrep} to search inside the app folders for the presence of specific keywords.

\subsubsection{Mining of Info from the Google Play Store} For all the packages extracted from the first mining from F-Droid, we performed an inspection to understand whether they were present also on the PlayStore. If so we gathered information about them from such platform. 

To find whether an app was released on the Play Store, we leveraged the particular format of the URL on the platform, http://play. google.com/store/apps/details?id=package\_name, where package \_name is used to uniquely identify the application. 

Using again the Java API of Selenium Chromedriver, we were able to mine the following information for each application: date of the last update, number of downloads, rating in stars (ranging from 1.0 to 5.0), number of ratings. %Those scraped metrics, of which we checked the correlation with the amount of Kotlin code with statistical tests, allowed us to answer RQ3. %We also gathered other information that were not usable for our statistics since they may vary according to the device: needed version of Android, number of version, size of the apk. 

During the analysis of the results of the scraping of the Play Store, it emerged that several applications were registered on F-Droid with package names of other Android apps already available on the Play Store. We hence manually checked the correspondence between the apps distributed through the official store and the package name declared in the F-Droid release. The comparison was based on the size of the apk, the title of the application (which is often different from the package name), and the application icon.

\subsubsection{GitHub Analysis and Mining} %We performed a two-fold analysis on GitHub, to (i) understand the trend of releases of Android apps featuring Kotlin code, and (ii) provide additional characterization to the apps mined from the F-Droid repositories whose source code is also released on GitHub.

%For the first analysis, we leveraged the GitHub APIs in a Python script, passing as parameters the Android keyword (that is searched in the name, description and readme files of the project) and Java and Kotlin as language parameter. The language parameter is not exclusive, i.e., a single repository can be considered of being written in multiple languages. As an additional parameter, we used the \emph{created} parameter, that indicates the date in which a given package was first uploaded on GitHub.

Since mining information of all the Android projects on GitHub was out of the scope of this experiment, we searched for an association between the projects mined from F-Droid and repositories hosted on GitHub. First, we created a script to find projects declaring in the manifest file the package names identifying the F-Droid projects. This search, however, led to many duplicates, because the same repository could be subject to clones or re-upload on GitHub. We hence applied another filtering phase to find the project that should likely correspond to the F-Droid one: we searched whether the "F-droid" keyword was present in the description or readme file of each of the GitHub repositories; if the search yielded a single result, we took that as the corresponding one and discarded all the others; if the search yielded multiple results, we resorted on taking the oldest repository (i.e., that with earliest creation date) based on the assumption that the others should likely be clones or forks of it. The correctness of the oldest repository was in all cases confirmed with a manual verification. 

For all the GitHub projects associated to those mined from F-Droid, we then performed a scraping to obtain popularity metrics; we checked the correlation of these metrics with the amount of Kotlin code in the repositories.% to give an answer to RQ3.

\subsubsection{Static Analysis of GitHub Repositories}

We cloned all the GitHub repositories of our context that featured Kotlin code, and examined the evolution of the repositories. We did so by using git commands inside a bash script. We examined the evolution of the repositories at release granularity, checking out all the tagged releases that could be extracted by using the \emph{git log} command.

\subsection{Analysis Method}

Starting from the set of all projects hosted on F-Droid, we defined four different sets of projects, subdivided based on the platforms where they were present: on F-Droid only, on F-Droid and PlayStore, on F-Droid and GitHub, on all the three platforms.

Since we wanted to characterize the diffusion of Kotlin, we restricted our repository to projects that were updated on F-Droid or GitHub after October 2017, i.e. since when Kotlin was officially supported by the Android Studio IDE. We have computed the diffusion metrics used to answer RQ1 on this subset of projects.

To count the lines of Kotlin code and compute the diffusion metrics, we have considered the most recently updated app package between the one contained in the .tar.gz mined from F-Droid and the package cloned from GitHub (if available).

By comparing the last update on the three repositories, we identified projects that were not kept up to date with respect to the correspondent projects on the other repositories. We defined a 45-days threshold to define a project as \emph{abandoned} on a given repository. We used the set of projects that were updated the last time after October 2017 and not abandoned on GitHub to analyze the history of Kotlin adoption, measuring the Evolution metrics and hence provididing our answer to RQ2.

To answer RQ3, we performed Wilkoxon Rank Sum tests to verify the existence of correlations between the presence of Kotlin in software projects, or the fact that a project has a majority of Kotlin code, with three different popularity metrics: the \emph{ratings} and \emph{downloads} on the Play Store, and the number of \emph{stars} on GitHub.

%0) filtraggio iniziale sulla qualità dei dati (?)
%1) dopo 2017 e perché (kotlin first-class dopo ottobre 2017)
%2) non abbandonati con definizione di abandoned
%3) su quali RQ*
%4) tipo di test statistico - Wilcoxon

\section{Results and Discussion}

\begin{table}
    \centering
    \small
        \caption{Considered sets of projects for context definition and Research Questions}

    \begin{tabular}{p{7cm}r}
    \toprule
    Project set & Size\\
    \midrule
        Total projects mined from F-Droid &  1860\\
        \quad - also on Play Store & 1013\\
        \quad - also on GitHub & 840 \\
    \midrule
    Projects updated after October 2017 (RQ1) & 1232\\
    \quad - with release history and not abandoned on GitHub (RQ2) & 145\\
   \quad - not abandoned on PlayStore (RQ3) & 475\\
    \bottomrule 
    \end{tabular}
    \label{tab:projectsets}
\end{table}

\begin{figure}
    \centering
    \includegraphics[width=\columnwidth]{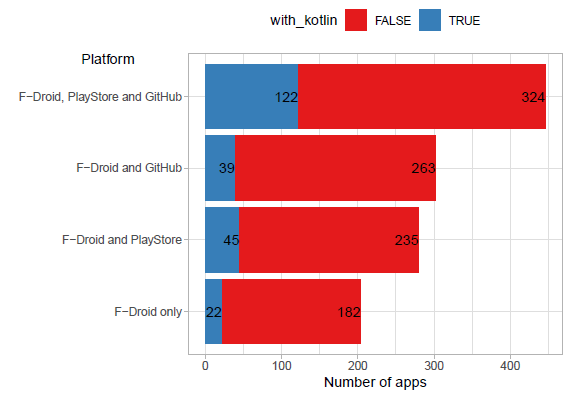}
    \caption{Number of projects on the considered platforms with last update after 2017}
    \label{fig:diffusion}
\end{figure}

\begin{table}
    \centering
    \caption{Statistics abount project abandonment (update on the store more than 45 days before the overall last update)}
    \small

    \begin{tabular}{lccc}
    \toprule
        &Ab. on & Ab. on & Ab. on\\
         &  F-Droid &  PlayStore &  GitHub\\
         \midrule
    F-Droid and PlayStore & 0.16 & 0.21 & -\\
    F-Droid and GitHub & 0.39 & - & 0.07\\
    F-Droid, PlayStore and GitHub & 0.44 & 0.45 & 0.05\\
    \bottomrule
    \end{tabular}
    \label{tab:abandonment}
\end{table}

Table \ref{tab:projectsets} reports the different sets of projects, and their size, that were used for the individual parts of the study. 

As of the beginning of May, 2019, when our final measurements were collected, we have mined a total of 1,860 projects from the F-Droid platform. 1,013 (53.51\%) of the OS projects on F-Droid were also released on the Google Play Store, and 840 (44.37\%) were also published as GitHub repositories. 476 (25.14\%) apps were available on all the three repositories.

1,232 projects have been updated after October 2017, i.e. since when Kotlin was adopted as a first-class programming language. The bar plot in figure \ref{fig:diffusion} reports the number of projects with and without Kotlin LOCs for all the subsets of projects that were updated after October 2017. The largest set was that of projects appearing on all the three repositories; on the other hand, just 204 of those projects (the 16.55\%) appeared on F-Droid only. This result suggests that Android OS developers rarely rely on the F-Droid repository only to publish their application.

By comparing the most recent update dates for projects that were present on multiple platforms, it emerged that many projects on F-Droid and on the PlayStore are not kept up to date with their GitHub counterparts (see table \ref{tab:abandonment}). Specifically, considering the projects that appear on all three repositories, near 50\% of latest releases on F-Droid and PlayStore are more than 45 days older than the latest tagged release on GitHub. This result was expected, since a 1-to-1 relationship between tagged releases on GitHub and releases on the stores is not likely. On the other hand, 5\% of the projects appearing on all repositories were no longer updated on GitHub: this subset of projects may indicate projects moved on other code hosting platforms or (if they are still updated on the PlayStore) turned closed-source.

%\begin{table}
%    \centering
%    \small
%        \caption{Diffusion metrics for projects with last update after 2017}
%
%    \begin{tabular}{lrrrrr}
%    \toprule
%    & Apps & $\overbar{KRL}$ & $\overbar{KRF}$ & KFPR &  KMPR\\
%    \midrule
%    F-Droid only & 204 & 0.07 & 0.07 & 0.11 & 0.07\\ 
%    F-Droid and PlayStore & 280 & 0.09 & 0.10 & 0.16 & 0.10\\
%    F-Droid and GitHub & 302 & 0.08 & 0.09 & 0.13 & 0.08\\
%    F-Droid, PlayStore and GitHub & 446 & 0.19 & 0.19 & 0.27 & 0.19\\
%    All platforms & 1232 & 0.12 & 0.13 & 0.19 & 0.12\\
%    \bottomrule
%    \end{tabular}
%    \label{tab:diffusion2}
%\end{table}
%
%\begin{table}
%\small
%    \centering
%        \caption{Diffusion metrics only for projects featuring Kotlin}
%    \begin{tabular}{lrrrr}
%    \toprule
%    &Apps & $\overbar{KRL}$ & $\overbar{KRF}$ & KMPR\\
%    \midrule
%    F-Droid only & 22 & 0.65 & 0.67 & 0.64\\
%    F-Droid and PlayStore & 45 & 0.57 & 0.62 & 0.60\\
%    F-Droid and GitHub & 39 & 0.65 & 0.68 & 0.64\\
%    F-Droid, PlayStore and GitHub & 122 & 0.68 & 0.71 & 0.69\\
%    All & 228 & 0.65 & 0.68 & 0.66\\
%    \bottomrule
%    \end{tabular}
%    \label{tab:diffusion2}
%\end{table}

\begin{table}
    \centering
    \small
        \caption{Diffusion metrics for projects with last update after 2017}

    \begin{tabular}{@{}lrrrrr@{}}
    \toprule
    &\multicolumn{2}{c}{All Apps}&\multicolumn{3}{c}{Kotlin Apps}\\
    \cmidrule(lr){2-3} \cmidrule(lr){4-6}
    & Apps & KFPR & $\overbar{KRL}$ & $\overbar{KRF}$ & KMPR\\
    \midrule
    F-Droid only & 204 & 0.11  & 0.65 & 0.67 & 0.64\\ 
    F-Droid and PlayStore & 280 & 0.16 & 0.57 & 0.62 & 0.60\\
    F-Droid and GitHub & 302 & 0.13 & 0.65 & 0.68 & 0.64\\
    F-Droid, PlayStore \& GitHub & 446 & 0.27  & 0.68 & 0.71 & 0.69 \\
    \addlinespace[0.5ex]
    All platforms & 1232  & 0.19 & 0.65 & 0.68 & 0.66 \\
    \bottomrule
    \end{tabular}
    \label{tab:diffusionAll}
\end{table}

%The results in table \ref{tab:diffusion2} show the measured Diffusion metrics on the set of projects that were updated after October 2017. It emerges that 19\% of the projects featured Kotlin LOCs, with 12\% of the projects having a majority of Kotlin LOCs on the total set of production code. On average, 12\% of the LOCs and 13\% of the code files of the mined apps are written with Kotlin.
%
%Analyzing the set of apps released on different sets of platforms, it can be seen that the amount of projects featuring Kotlin (KFPR) is typically increased if they are released also on GitHub, and that the apps released only on F-Droid featured the lowest relative amount of Kotlin code. This result can be paired with the abandonment frequence reported in table \ref{tab:abandonment}, and suggests -- as reasonable -- that the transition to Kotlin is rather gradual and typically involves the GitHub repository first, and the stable releases on the app markets later. 
%
%
%Results in table \ref{tab:diffusion2} show the diffusion metrics average metrics measured only on the 228 projects that feature Kotlin. These results show that, when Kotlin is adopted, on average the majority of the code of an application (65\% of LOCs and 68\% of files) is written with it. This result may suggest that the Kotlin language is preferred to Java when adopted, or that the development guidelines of Android apps encourage a full conversion of Android projects to Kotlin rather than a coexistence with existing Java code.

The results in table \ref{tab:diffusionAll} show the measured Diffusion metrics on the set of projects that were updated after October 2017. It emerges that 19\% of the projects featured Kotlin code.

Analyzing the set of apps released on different sets of platforms, it can be seen that the amount of projects featuring Kotlin (KFPR) is typically increased if they are released also on GitHub, and that the apps released only on F-Droid featured the lowest relative amount of Kotlin code. This result can be paired with the abandonment frequency reported in table \ref{tab:abandonment}, and suggests -- as one could reasonably expect -- that the transition to Kotlin is rather gradual and typically involves the GitHub repository first, and the stable releases on the app markets later. 

The three rightmost columns in table \ref{tab:diffusionAll} report the average metrics measured only on the 228 projects that feature Kotlin. These results show that, when Kotlin is adopted, on average the majority of the code of an application (65\% of LOCs and 68\% of files) is written with it. This result may suggest that the Kotlin language is preferred to Java when adopted, or that the development guidelines of Android apps encourage a full conversion of Android projects to Kotlin rather than a coexistence with existing Java code.

\vspace{1ex}
\noindent\fbox{%
    \parbox{0.97\columnwidth}{%
        \textbf{Answer to RQ1}: Nearly one fifth of the 1232 mined from F-Droid that were updated after October 2017 featured Kotlin code, with 2/3 of those projects featuring a majority of Kotlin code. The diffusion of Kotlin code in Android application largely increases when looking at the subset of apps whose source code has been also released on GitHub.
    }%
}\\[2ex]

\begin{figure}
    \centering
    \includegraphics[width=\columnwidth]{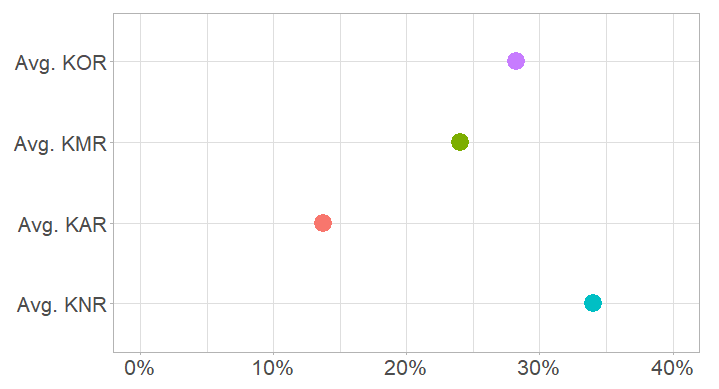}
    \caption{Average evolution metrics for apps released on GitHub (set of 145 projects)}
    \label{fig:evolution1}
\end{figure}

\begin{figure}
\centering
\includegraphics[width=\columnwidth]{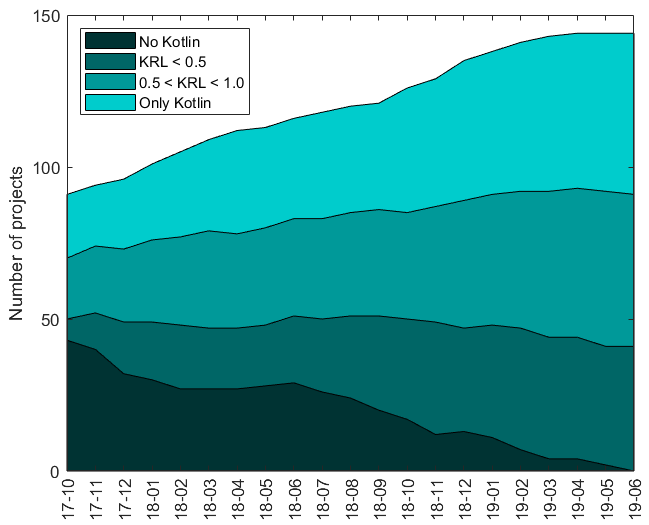}
\caption{Evolution of the KRL metric on the lifespan of GitHub projects featuring Kotlin in the latest update (set of 145 projects)}
\label{fig:evolution2}
\end{figure}

The history of the Kotlin adoption was evaluated on projects that had their last update after October 2017, that were not abandoned on GitHub, and that featured at least a single tagged release (in addition to the master branch considered for the static analysis of source code). This last conditions allowed a release-by-release comparisons to between the code of consecutive releases. By applying this filtering procedure, we came up with 145 projects to analyze. The analyzed projects had an average lifespan (between the first tagged release and the last update) of 862 days, corresponding to an average amount of 34.3 releases. The lifespans were rather variable, ranging from 5 to 3458 days (2 to 308 releases).

Figure \ref{fig:evolution1} reports the average Evolution metrics measured on this set of projects. On average, 30\% of tagged releases of the projects (corresponding to around 10 releases, and 258 days) did not feature any Kotlin LOC. Then, the projects experienced quite a fast transition to a relevant adoption of Kotlin: the amount of releases with 0-50\% Kotlin code (avg. Kar in the graph) was, in fact, just 13.7\% of the average lifespan; on the other hand, 28.2\% of the average lifespan featured only Kotlin code.

Figure \ref{fig:evolution2} shows the trend of Kotlin adoption by month since October 2017, on the set of 145 projects that featured Kotlin on last GitHub update. With few exceptions in the first months of 2018, the trends of all the considered variables confirm that the relative amount of Kotlin code constantly grew with time in the analyzed projects, while Java code was gradually discarded. Since the beginning of the considered period, more than half of the projects already featured a majority of Kotlin code, suggesting that the transition to Kotlin was common practice among Android developers already before it becoming first-class programming language for the domain.\smallskip

\vspace{1ex}
\noindent\fbox{%
    \parbox{0.97\columnwidth}{%
        \textbf{Answer to RQ2}: Most of the projects that featured Kotlin in their latest release showed a quick transition from Java to Kotlin during their lifespan. Specifically, on average, 70\% of releases featured Kotlin, and in 30\% releases Kotlin code no longer coexisted with Java code.
    }%
}\\[2ex]

\begin{table}
\small
    \centering
    \caption{Null Hypotheses and Wilcoxon test results}

    \begin{tabular}{@{}l>{\raggedright}m{4.5cm}cc@{}}
    \toprule
        Name & Description & p-value & Decision \\
        \midrule
         $Hrk_{0}$ & Using Java or Kotlin has no impact on the average Play Store ratings of an app. & 0.633 & No-Reject \\
         $Hdk_{0}$ & Using Java or Kotlin has no impact on the number of downloads from the Play Store of an app. & 0.0666 & No-Reject\\

$Hsk_{0}$ & Using Java or Kotlin has no impact on the number of GitHub stars for an app. & 0.0002 & \textbf{Reject}\\

$Hrm_{0}$ & The relative amount of Kotlin code has no impact on the average Play Store ratings of an app. & 0.687 & No-Reject\\

$Hdm_{0}$ & The relative amount of Kotlin code has no impact on the number of downloads from the Play Store of an app. & 0.0866 & No-Reject\\

$Hsm_{0}$ & The relative amount of Kotlin code has no impact on the number of GitHub stars of an app. & 0.560 & No-Reject\\

         \bottomrule
    \end{tabular}
    \label{tab:wilcoxon_likert}
\end{table}

%table for hypotheses

\begin{figure}
\centering
\begin{subfigure}{1\columnwidth}
   \includegraphics[width=1\columnwidth]{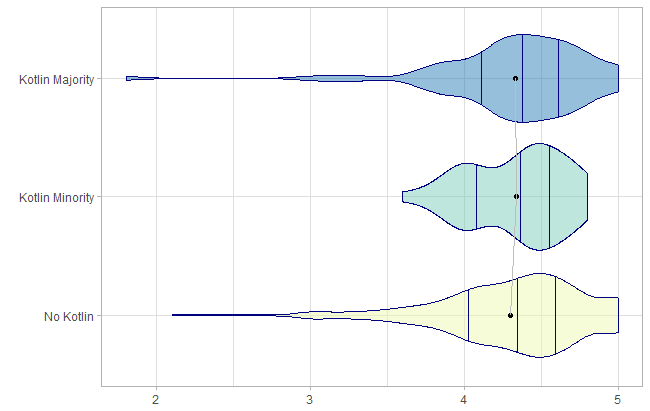}
   \caption{Average rating on the Play Store}
   \label{fig:rq3_1} 
\end{subfigure}
\par\bigskip
\begin{subfigure}{1\columnwidth}
   \includegraphics[width=1\columnwidth]{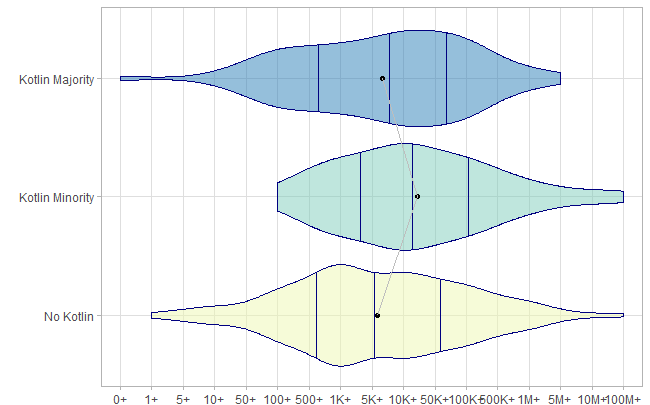}
   \caption{Number of downloads from the Play Store}
   \label{fig:rq3_2}
\end{subfigure}
\par\bigskip
\begin{subfigure}[b]{0.95\columnwidth}
   \includegraphics[width=1\columnwidth]{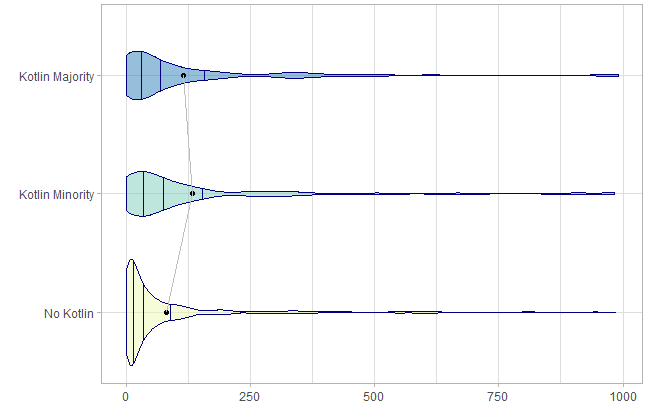}
   \caption{Number of Stars of the GitHub project}
   \label{fig:rq3_3}
\end{subfigure}

\caption{Distribution of popularity metrics for apps updated after October 2017 (statistics on a set of 475 projects)}
\label{fig:distributions}
\end{figure}

To find whether the adoption of Kotlin has any impact on the perception of the projects by end users or other developers, we sought for correlations between the presence of Kotlin code and the following popularity metrics: the average user \emph{Rating} and the number of \emph{Downloads} on Play Store, and the number of \emph{Stars} on GitHub, i.e. the number of developers that marked the project as a favorite. The analysis of popularity metrics is crucial for developers, that need to tailor their development based on several sources of feedback \cite{nayebi2018app}. We performed the analysis only on projects that were not abandoned, according to our definition, on the platforms where the respective popularity metrics were gathered.

Table \ref{tab:wilcoxon_likert} reports the definition of the null hypotheses and the resulting p-values of the Wilkoxon Paired Signed Rank tests that we applied to the distributions of the measured popularity metrics. Figures \ref{fig:rq3_1} to \ref{fig:rq3_3} show, respectively, the distributions of ratings, downloads and stars, measured for projects that had no Kotlin, for projects that featured Kotlin, and for projects that featured a majority of Kotlin code. 

We can observe that some distributions were quite different, especially those regarding the number of stars on GitHub (fig. \ref{fig:rq3_3}). In fact, $Hsk_0$ is the only null hypothesis that we could reject (p-value of the Wilkoxon test equal to 0.0002), meaning that there is statistical evidence that the use of Kotlin in a software project has an influence on the number of GitHub stars for an app. This result can be justified by considering the novelty of the Kotlin programming language, that may push other developers to follow existing projects with Kotlin to inspect the language and its peculiarities. On the other hand, $Hsm_0$ could not be rejected, hence we spotted no statistically significant influence of a majority of Kotlin code on the number of stars of a GitHub repository.

The distributions of the average rating on the Play Store (fig. \ref{fig:rq3_1}) showed a substantial similarity, confirmed by the relative Wilkoxon tests: hence, we could not reject $Hrk_0$ and $H_rm0$. The applications that featured Kotlin code, however, showed a slightly higher average rating than the apps developed with Java only.

The distributions of the number of downloads from the Play Store (fig. \ref{fig:rq3_2}) showcased an higher average value for projects featuring Kotlin code. However, the influence did not prove statistically significant (p-value of the Wilkoxon test equal to $0.06.6$).

\vspace{2ex}
\noindent\fbox{%
    \parbox{0.97\columnwidth}{%
        \textbf{Answer to RQ3}: The presence of Kotlin code had a statistically significant influence on the number of stars obtained by Android repositories on Google. In general, the Kotlin featuring Apps had, on average, higher popularity metrics than those developed with Java only. Such differences, howeverm did not prove statistically significant.
    }%
}\\[2ex]

\section{Threats to Validity}

\emph{Internal validity} threats concern factors that may affect a dependent variable. To compute the diffusion metrics and to obtain the subset of projects to answer RQ2, we labeled the projects as featuring Kotlin analyzing only the latest available version of the code. This filtering phase would exclude projects that no longer contain Kotlin, and hence add biases to our results. We also defined a project as \emph{abandoned} on a platform if the last update on that platform dated more than 45 days before the latest update on all repositories. Different choices for such threshold may lead to significantly different results for the considered evolution metrics.

\emph{External validity} threats concern the generalization of the results. Our findings are based on packages mined from the F-Droid app market and on their counterparts released on the Play Store and on GitHub. It is not assured that the computed metrics and the discussion based on them are applicable to other sets of applications. The findings are also limited to the usage of Kotlin in mobile apps, thus are not generalizable to other domains where the language can be adopted as an alternative to Java.

\emph{Conclusion validity} threats concern drawing the appropriate conclusion based on test results. In our hypothesis testing we adopted non-parametric tests to account for the non-continuous and non-normal distribution of the variables. We draw our conclusions using the customary 5\% type I error threshold. 

It is also worth highlighting that the detected correlation between the presence of Kotlin in source code of the apps and the number of GitHub stars does not imply causality. It is therefore possible that the most active projects, that already had a high number of stars on the platform, have then moved to Kotlin, and not the vice-versa, i.e. that projects with Kotlin attracted more stars.

\section{Related Work}

Many works in literature performed mining of app markets to analyze programming trends and characteristics of Android applications. Several studies were based on the F-Droid repository: Grano et al. studied software evolution and quality improvement of Android code \cite{grano2017android}; Kochhar et al. analyzed the adoption of test code to understand the testing culture of mobile app developers \cite{kochhar2015understanding}; Freiling et al. analyzed the apks to gather information about the code obfuscation techniques used by Android developers \cite{freiling2014empirical}; Cruz et al. analyzed the correlation between the adoption of testing techniques on Android applications and various popularity metrics on GitHub and on the PlayStore \cite{Cruz2019}. Other mining studies involved apks mined from the Google PlayStore directly: Munaiah et al. analyzed more than 60 thousand applications to classify them as malicious or benign \cite{munaiah2016darwin}; Avdiienko et al. analyzed applications to spot abnormal usage of sensitive data \cite{avdiienko2015mining}; many studies focused on opinion mining from Play Store Reviews: Genc et al. provide a systematic literature review of works leveraging such practice \cite{genc2017systematic}. Other large-scale studies were based on all the Android applications whose source code was published on GitHub: in our previous work, we analyzed the scripted testing diffusion and evolution on a data set of 280,000 Android projects \cite{Coppola2019}.

Quite a few empirical studies have focused on Kotlin since the spawning of the programming language. Shah et al. investigated the code obfuscation on applications developed with Kotlin \cite{kt1}. Several studies aimed at comparing Kotlin code with Java code: Flauzino et al. analyzed more than 6 millions lines of code, concluding that on average Kotlin code shows fewer smells \cite{kt2}; Banerjee et al. concluded that Kotlin code is more concise and safer than Java code \cite{banerjee2018comparative}.

\section{Conclusion and future work}

%RESUME

Among the several languages proposed for developing native mobile apps, Kotlin is currently gaining attention among developers, and is affirming as one fo the main alternatives to Java. The goal of the empirical study described in this manuscript was to provide an insight on the current level of penetration of the language in open-source Android projects. To do so, we mined all projects hosted on F-Droid and cloned -- if existing -- the correspondent repositories on the GitHub platform, to find their most recently updated source code. After restricting our analysis at projects updated after October 2017 (when Android Studio 3.0, officially adopting Kotlin, was released), we statically analyzed their source code and its evolution, we computed a set of metrics we definition in orderd to quantify the transition to Kotlin. We also collected popularity metrics for the analyzed projects, from both GitHub and Google Play Store -- when apps were also released on the market --. 

On our final set of 1232 applications, we found that 19\% of projects featured Kotlin code. Among those projects the transition from Java to Kotlin was most of the times fast and unidirectional: the ratio of Kotlin over total code on those projects was, with few exception, always increasing during their evolution from October 2017 until the moment we carried the measurement, in early May 2019. Projects with Kotlin exhibited, on average, higher values for the popularity metrics that we considered, namely the rating and the number of downloads on the PlayStore, and the number of stars on GitHub. More specifically, we found a statistically significant influence on the latter metric.

 These results seemingly confirm the selling points of Kotlin vs. Java coding that have been put in light by comparison studies already available in literature.
 
%RESULTS, A CHI SERVONO, ACTIONABILITY

Our results can provide useful evidence for Android applications developers:

\begin{itemize}
    \item Developers not familiar with Kotlin might consider worthwile learning it as a new language since one out of four open source applications on the F-Droid market is written in Kotlin;
    \item Leaders of open source projects interested in migrating their applications to Kotlin, should know that -- when adopted -- the language appears to be preferred to Java by fellow developers;
    \item When considering the alternatives among staying with Java, using Kotlin only in part, or using mostly Kotlin, developers should know that no significant impact was observed on user's appreciation of the app.
\end{itemize}

 The information about projects featuring Kotlin released on the F-Droid platform, that we published on our website\footnote{http://softeng.polito.it/coppola/wama\_paper\_data.csv}, may be useful for researchers aiming at investigating more properties of the Kotlin language, and performing additional comparisons with Java.

%FUTURE WORK

As our future work, we plan to extend the analysis on GitHub projects to all \emph{Android} projects featuring Kotlin. A large-scale inspection on all projects with Kotlin cloned from GitHub would give a valuable perspective on the current adoption of the Kotlin programming language. Additionally, we plan to analyze the programming patterns adopted by developers using Kotlin in the Android environment, and to further inspect Kotlin apps for what concerns vulnerabilities, testing and bug. Those analyses may provide valuable insights and possible guidelines for better development with the Kotlin language.

\bibliographystyle{unsrtnat}
\bibliography{sample-bibliography}

\balance

\end{document}